\documentclass[11pt, a4paper]{article}

\usepackage[a4paper, total={16cm, 25cm}]{geometry}
\usepackage{amsmath}
\usepackage{graphicx}
\usepackage{natbib}
\usepackage{amsthm}
\usepackage{color}
\usepackage{footmisc}
\usepackage{url}

\usepackage[frozencache]{minted}
\usepackage{listings}
\usepackage{xcolor}
\usepackage{setspace}
\usepackage{caption}
\usepackage{booktabs}
\usepackage{siunitx}
\usepackage{booktabs}

\newtheorem{rational for conjecture}{Rational for Conjecture}

\usepackage{algorithm}
\usepackage{algorithmic}

\lstdefinestyle{mystyle}{
    backgroundcolor=\color{lightgray!20},   
    frame=tb,
    framesep=2mm,
    basicstyle=\setstretch{1.2}\small\ttfamily, 
    breaklines=true,
    breakatwhitespace=true,
    breakindent=0pt,
    captionpos=b,
    keepspaces=true,
    showstringspaces=false
}

\begin{document}

\LARGE
\begin{center}
  \textbf{Using Large Language Models to Suggest Informative Prior Distributions in Bayesian Statistics} \\[10mm]
\end{center}

\normalsize

\begin{center}
\begin{tabular}{cc}
Michael A. Riegler & Kristoffer Herland Hellton \\
\small Simula Research Laboratory, Oslo, Norway & \small OsloMet, Oslo, Norway \\
& \\
Vajira Thambawita & Hugo L. Hammer \\
\small SimulaMet, Oslo, Norway & \small OsloMet and SimulaMet, Oslo, Norway \\
\end{tabular}
\end{center}

\begin{abstract}
Selecting prior distributions in Bayesian statistics is a challenging task. Even if knowledge already exists, gathering this information and translating it into informative prior distributions is both resource-demanding and difficult to perform objectively.

In this paper, we analyze the idea of using large-language models (LLMs) to suggest suitable prior distributions. The substantial amount of information absorbed by LLMs they have a potential for suggesting knowledge-based and more objective informative priors. We have developed an extensive prompt to not only ask LLMs to suggest suitable prior distributions based on their knowledge but also to verify and reflect on their choices.

We evaluated the three popular LLMs Claude Opus, Gemini 2.5 pro, and ChatGPT o4-mini for two different real datasets: an analysis of heart disease risk and an analysis of variables affecting the strength of concrete. For all the variables, the LLMs were capable of suggesting the correct direction for the different associations, e.g., that the risk of heart disease is higher for males than females or that the strength of concrete is reduced with the amount of water added. The LLMs suggested both moderately and weakly informative priors, and the moderate priors were in many cases too confident, resulting in prior distributions with little agreement with the data. The quality of the suggested prior distributions was measured by computing the distance to the distribution of the maximum likelihood estimator (``data distribution") using the Kullback-Leibler divergence. In both experiments, Claude and Gemini provided better prior distributions than ChatGPT. For weakly informative priors, ChatGPT and Gemini defaulted to a mean of 0, which was "unnecessarily vague" given their demonstrated knowledge. In contrast, Claude did not. This is a significant performance difference and a key advantage for Claude's approach.

The ability of LLMs to suggest the correct direction for different associations demonstrates a great potential for LLMs as an efficient and objective method to develop informative prior distributions. However, a significant challenge remains in calibrating the width of these priors, as the LLMs demonstrated a tendency towards both overconfidence and underconfidence.

A link to our code will be made available in the published version.
\end{abstract}

\section{Introduction}

\label{sec:introduction}

Large Language Models (LLMs) can generate human-like text with remarkable fluency and coherence. Their learned parameters have also absorbed a substantial amount of knowledge, making them useful for many real-world tasks. Examples include clinical decision support and patient communication within healthcare~\cite{li2024revolutionizing, busch2025current}, fraud detection and market sentiment analysis within finance~\cite{nie2024survey, esma2024leveraging}, and enabling personalized learning and real-time student support within education~\cite{wang2024large, alhafni2024llms}, to name a few. However, the knowledge LLMs contain is inherently unstructured and derived from patterns in their training data. This can lead to outputs that are inconsistent, biased, or factually incorrect—a phenomenon often termed "hallucination"~\citep{ji2023survey}. Therefore, while LLMs are useful sources for quickly accessing information, relying on them in isolation carries significant risks.

Empirical reasoning is a fundamental part of scientific discovery and knowledge acquisition in general. Statistical modeling and analysis are the formal language of this process, allowing researchers to draw robust conclusions and quantify uncertainty. In many situations, it can be beneficial to combine pre-existing knowledge or beliefs with information from empirical data to leverage both sources of information~\citep{gelman2013}. Bayesian statistics offers a popular framework for this, consisting of a likelihood function, which represents the information from the empirical data, and prior distributions, which are formulated to represent pre-existing knowledge and beliefs. These two sources of information are combined using Bayes' rule, resulting in the posterior distribution. However, formulating informative prior distributions is often difficult, and practitioners frequently resort to using wide, non-informative priors \cite{mikkola2024prior}. This practice means that existing knowledge is not included in the analysis, leading to a potential loss of valuable information. While one could conduct a substantial literature review for a given problem to develop informative priors, this approach is highly resource-demanding and challenging to compile into objective and informative prior distributions~\citep{gelman2013}.

Given the vast amount of knowledge absorbed by LLMs, an appealing idea is to combine this knowledge with the information in available empirical data. However, how to perform such a combination in practice is not obvious. In this paper, we analyse the idea of casting this problem into a Bayesian framework, where we let LLMs represent pre-existing knowledge in the form of informative prior distributions. To develop suitable LLM-based prior distributions, we describe the problem we want to study to the LLM—for instance, potential risk factors for a disease—and specify the Bayesian model we will use to analyze the data, such as a logistic regression model with Gaussian prior distributions for the regression coefficients. We then ask the LLM to use its knowledge to suggest suitable values for the hyperparameters of these prior distributions. We require the model to provide a detailed justification for the value selected for each hyperparameter. We also ask the LLM to suggest multiple sets of priors (e.g., moderately and weakly informative) based on its domain knowledge and to assign a relative weighting or confidence score to each suggested set. The exact prompts used in this study are available, along with the code, via the GitHub link provided.

Previous research has explored the idea of using Large Language Model (LLM) knowledge to improve models through methods such as feature selection and engineering \cite{hollmann2025accurate}. For instance, \cite{requeima2024llm} use LLMs to incorporate user's prior knowledge, while \cite{capstick2024using} leverage them to suggest prior distributions for improving the prediction of urinary tract infections in people living with dementia. Others have used LLMs to generate synthetic data from public datasets, which define priors on linear models through a separate likelihood term \cite{gouk2024automated}, or to search over a potentially infinite set of concepts within Concept Bottleneck Models \cite{feng2024bayesian}. In contrast to these works, our focus is on analyzing how the suggested prior distributions agree with the data and developing effective prompts for LLMs to obtain well-formed prior distributions.

The main contributions of this paper are:
\begin{itemize}
\item We develop an extensive prompt for LLMs to suggest suitable prior distributions, requiring the LLM to reflect on its use of knowledge, propose multiple prior distribution sets, and critically justify its choices with confidence scores. This approach is a key contribution because it elevates the interaction from a simple query to a structured, interpretable knowledge elicitation process that produces more reliable outputs.
\item We visualize and systematically quantify the quality of the prior distributions suggested by the three popular LLMs. We find that the more confident "moderately informative" priors were often worse (had higher KL divergence) than the weakly informative ones. It suggests a meta-level overconfidence: the LLMs are not only overconfident in their parameter estimates but also in their assessment of which set of priors is better.
\item We evaluate whether the inclusion of LLM-based prior information can improve prediction performance.
\end{itemize}

\section{Methodology}

Let the stochastic variable $\mathcal{D}$ denote a dataset. For instance, $\mathcal{D}$ may consist of $p$ input features $\mathbf{x}i = x_{i1}, \ldots, x_{ip}$ for $i = 1, \ldots, n$ samples, along with their associated responses $Y_i$, for $i = 1, \ldots, n$. The goal of Bayesian statistics is to infer unknown parameters $\theta$, for example regression coefficients, by incorporating both prior knowledge and the information contained in the data $\mathcal{D}$. Prior knowledge and beliefs are captured by the prior distribution $p(\theta)$. The likelihood function $p(\mathcal{D}|\theta)$ expresses the probability of observing the data $\mathcal{D}$ given a specific set of parameter values $\theta$.

\begin{figure}
\centering
\includegraphics[width=\linewidth]{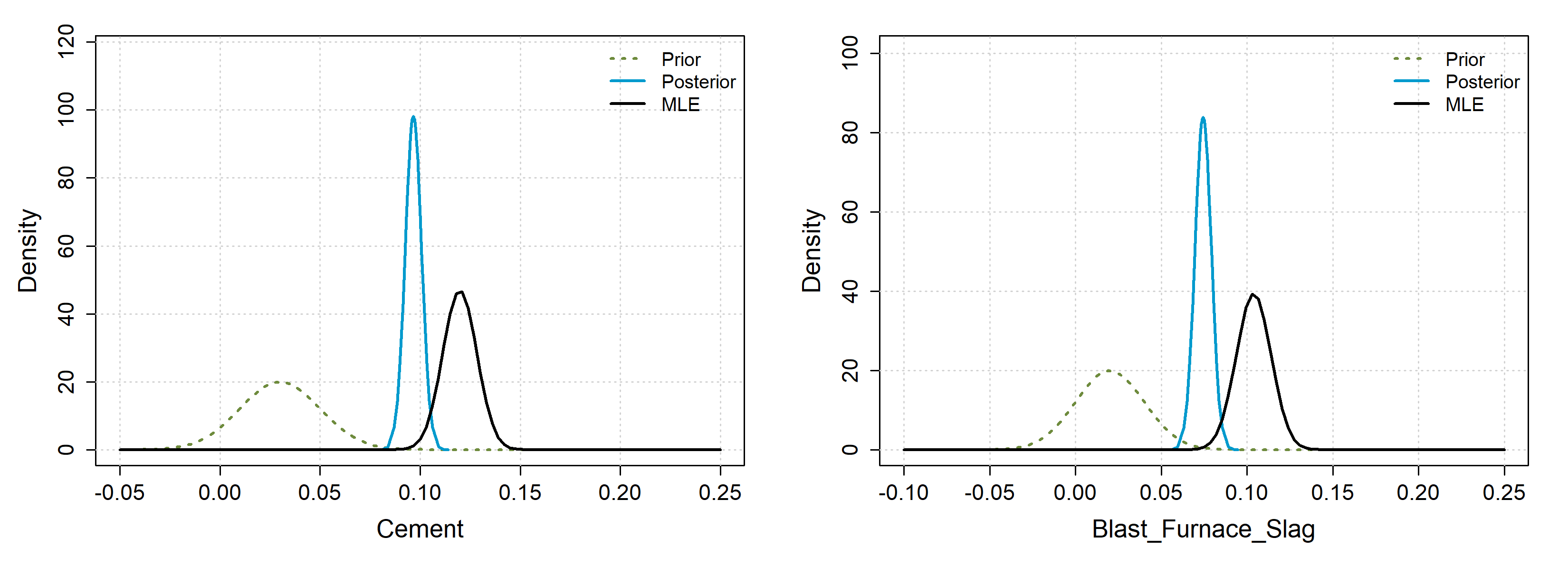}
\caption{Cement strength example: Visualization of prior, MLE, and posterior distributions. The prior distributions were suggested by ChatGPT.}
\label{fig:5}
\end{figure}

A common and intuitive approach to parameter inference is to find the value of $\theta$ that maximizes the likelihood function. This is known as the maximum likelihood estimator (MLE). Since the MLE is a function of the data $\mathcal{D}$, for example, the sample mean, it is also a stochastic variable that follows a distribution. In Bayesian statistics, by contrast, inference about $\theta$ is based on evaluating the posterior distribution:
\begin{equation}
p(\theta|\mathcal{D}) = \frac{p(\mathcal{D}|\theta) p(\theta)}{p(\mathcal{D})} \propto p(\mathcal{D}|\theta) p(\theta)
\end{equation}
A key distinction in Bayesian statistics is that the unknown parameter is treated as a stochastic variable, while in frequentist statistics it is considered fixed. The posterior distribution combines the information regarding $\theta$ from both the data and the prior distribution. Figure \ref{fig:5} illustrates two examples from the cement strength experiment in Section \ref{sec:concrete}, where the plots show the prior, MLE, and posterior distributions for the regression coefficients of the variables cement and blast furnace slag. The resulting posterior distributions lie between the prior and MLE distributions. Furthermore, we observe that there is little agreement between the data (represented by the MLE) and the prior distribution, which is generally undesirable. This issue will be discussed further in Section \ref{sec:evaluation}.

As described in the introduction, the aim of this paper is to evaluate the potential of state-of-the-art large language models (LLMs) to suggest informative and reliable prior distributions. This enables the use of both the knowledge encoded in LLMs and the observed data as sources of information. In Section \ref{sec:prompts}, we describe the prompts sent to the LLMs, and in Section \ref{sec:evaluation}, we outline how we evaluate the quality of the suggested prior distributions.

\subsection{LLM Prompts}
\label{sec:prompts}

Listing~\ref{lst:1} shows the main components of the suggested prompt, illustrated using the heart disease example from Section \ref{sec:heart}. The complete prompts used in this study are available alongside the code via the link provided at the end of the abstract. The objective is not only to obtain prior distributions from the LLMs, but also to require them to explain their reasoning, propose multiple sets of priors, and discuss their strengths, weaknesses, and level of confidence. 
This structured elicitation process is a core part of our methodology; it is designed to move beyond a simple query that might yield an unsubstantiated number, and instead forces the LLM into a more rigorous process of justification. This makes the resulting suggestions more reliable and interpretable than those from a simple 'ask and receive' approach. This prompt was not the first one used but was rather a result of iterative refinement that can provide valuable insights for other researchers looking to use LLMs for similar tasks.
\begin{listing}[H]
\begin{lstlisting}[style=mystyle]
You are an expert in biostatistics and cardiovascular epidemiology. For the logistic regression model provided below, which predicts coronary artery disease (CAD), your task is to propose and justify suitable normally distributed prior distributions for all regression parameters ($\beta_0, \ldots, \beta_6$).

**Model Details:**
Response: $y \in \{0,1\}$ (0 = healthy, 1 = CAD)
Linear Predictor: $\eta = \beta_0 + \beta_1 \text{age} + \beta_2 \text{sex} + \beta_3 \text{trestbps} + \beta_4 \text{chol} + \beta_5 \text{thalach} + \beta_6 \text{oldpeak}$
Predictor Details:
    * `age`: in years
    * `sex`: categorical (1 = male; 0 = female)
    * `trestbps`: resting blood pressure (in mm Hg on admission to the hospital)
    * `chol`: serum cholesterol in mg/dl
    * `thalach`: maximum heart rate achieved (beats per minute)
    * `oldpeak`: ST depression induced by exercise relative to rest (mm)

**Your Response Should:**
1. **Leverage Knowledge & Simulate Tool Use**: Briefly state how you'll use your existing knowledge of CAD risk factors and logistic regression modelling (simulating the consultation of relevant literature or databases for effect sizes and typical parameter ranges) to inform your suggestions.
2. **Propose Multiple Prior Sets**: Generate at least two distinct sets of prior distributions (e.g., "Suggestion A: Moderately Informative Priors based on Domain Knowledge" and "Suggestion B: Weakly Informative / More Conservative Priors").
3. **Detailed Justification for Each Parameter**: For each parameter ($\beta_0$ through $\beta_6$) within each suggested set, provide:
    * The specific normal prior distribution: $N(\text{mean}, \text{standard deviation}^2)$. Clearly state the mean and standard deviation.
    * ... (not inclded for the sake of brevity)
4. **Comparative Evaluation & Weighting**:
    * Critically evaluate and compare the different sets of priors you've proposed. ... (not inclded for the sake of brevity)
    * Assign a relative weighting or confidence score (e.g., Suggestion A: 60%, Suggestion B: 40%) to your proposed sets, and clearly explain the reasoning behind this weighting. ... (not inclded for the sake of brevity).
\end{lstlisting}
\caption{Main components for the prompt used in the heart disease example in Section \ref{sec:heart}.}
\label{lst:1}
\end{listing}

\subsection{Evaluation of LLM Priors}
\label{sec:evaluation}

In this section, we describe how we evaluate the quality of prior distributions suggested by LLMs. On the one hand, we require that the prior distribution is not in significant disagreement with the observed data, that is, the data should not be too ``surprising'' under the prior. In the examples in Figure~\ref{fig:5}, the observed data are clearly a substantial surprise to the prior distributions. To avoid such surprises, it is common to select wide or uninformative priors. On the other hand, we also want the priors to be informative and contribute useful information to the analysis. Simply put, selecting a good prior involves a trade-off between reducing the risk of surprise and being informative.

A fundamental method for assessing how surprising the data are under the prior is to generate synthetic data, $\mathcal{D}^*$, from the marginal data distribution:
\begin{equation}
\mathcal{D}^* \sim p(\mathcal{D}^*) = \int p(\mathcal{D}^*|\theta) p(\theta) \, \text{d} \theta
\end{equation}
also referred to as the prior predictive distribution~\cite{evans2006checking}. The degree of disagreement between the synthetic data $\mathcal{D}^*$ and the observed data $\mathcal{D}$ is then quantified using summary statistics such as the mean, standard deviation, and quantiles. A drawback of this method is that it can be challenging to identify suitable statistics, especially for high-dimensional data, to meaningfully capture the level of disagreement.

In this paper, we instead focus on directly comparing the two sources of information: the likelihood and the prior distribution, a comparison that has received less attention in the literature most probably because the MLE is more of a frequentist than Bayesian concept. While the exact distribution of the MLE is usually unknown, it can often be efficiently estimated from the data. For example, it is well known that the MLE is asymptotically Gaussian~\cite{casella2024statistical}. Non-parametric approaches, such as bootstrapping, can also be used for estimation.

When assessing the quality of a prior distribution, we aim to evaluate both how surprising the data are under the prior and how informative the prior is. The Kullback–Leibler (KL) divergence is a metric well suited for this purpose:
\begin{equation}
\label{eq:1}
D_{\mathrm{KL}}(p || \hat{p}_{\text{MLE}}) = \int \hat{p}_{\text{MLE}}(\theta) \log \frac{\hat{p}_{\text{MLE}}(\theta)}{p(\theta)} , \text{d} \theta,
\end{equation}
where $\hat{p}_{\text{MLE}}(\theta)$ denotes the approximation of the MLE distribution based on the observed data. The intuition is that if the prior assigns low probability mass to regions where $\hat{p}_{\text{MLE}}(\theta)$ is high, then the KL divergence will be large, as the prior appears to be surprised by the data. Since $p(\theta)$ appears in the denominator, the divergence substantially penalizes prior distributions that are in conflict with the data. In contrast, if the MLE is a surprise to the prior, the penalty is smaller, an asymmetry that aligns well with practical reasoning.

The KL divergence has previously been used to measure prior–data conflict, e.g., in Nott et al.~\cite{nott2020checking}, but typically with the focus on comparing the prior with the posterior. However, in our experiments we observed that in many cases where the MLE was a substantial surprise to the prior distribution, it could still be a substantial overlap between the prior and the posterior. Therefore we find it more useful to compare the prior directly with the MLE distribution.

Finally it is important to remark that in a pure Bayesian framework, a prior distribution is not inherently 'wrong' simply because it diverges from the observed data. For the purpose of evaluating the quality of the suggested LLM priors, we utilize the MLE distribution as a practical, data-driven benchmark even if it is not so common to include in Bayesian analysis.

\section{Experiments and Results}

To evaluate the potential for LLMs to suggest useful informative prior distributions, we used three state-of-the-art models: ChatGPT-4o-mini \cite{openai_chatgpt-4o-mini_2024}, Gemini 2.5 Pro \cite{google_gemini_1.5_pro_2024}, and Claude Opus \cite{anthropic_claude_3_opus_2024}.

\subsection{Heart Disease}
\label{sec:heart}

In this example, we consider the Cleveland Heart Disease dataset~\cite{heart_disease_45}. The aim is to analyse the association between coronary artery disease (CAD) and the predictors shown in Table \ref{tab:1}. CAD is a binary variable, where 0 indicates a healthy individual and 1 indicates the presence of CAD. We assume a logistic regression model for the problem, with CAD as the response and the variables in Table \ref{tab:1} as the predictors. We further assume Gaussian prior distributions for the regression coefficients.
\begin{table}
\centering
\caption{Description of Variables.}
\label{tab:1}
\begin{tabular}{@{}llc@{}}
\toprule
\textbf{Variable} & \textbf{Description} & \textbf{Unit} \\
\midrule
age & Age of the individual & years \\
sex & Biological sex (1 = male; 0 = female) & categorical \\
trestbps & Resting blood pressure on admission & \si{\mmHg} \\
chol & Serum cholesterol level & \si{mg/dl} \\
thalach & Maximum heart rate achieved & beats per minute \\
oldpeak & ST depression induced by exercise relative to rest & mm \\
\bottomrule
\end{tabular}
\end{table}

When running the prompt as shown in Listing \ref{lst:1}, all three LLMs responded convincingly, referring to reliable sources such as the Framingham Heart Study~\cite{mahmood2014framingham}, the MONICA project~\cite{monica_monograph_2003}, and various meta-analyses. All the LLMs were able to recall that the regression coefficients in logistic regression are represented by log-odds ratios. This required the LLMs to convert any known linear associations into the log-odds format. All of the LLMs successfully performed this conversion. The LLMs suggested two sets of priors: moderately and weakly informative. ChatGPT was 60\% and 40\% confident in the moderately and weakly informative priors, respectively, while Gemini and Claude were 65\% and 35\% confident, respectively. In other words, all three LLMs expressed the highest confidence in the moderately informative priors.

\begin{figure}
\centering
\includegraphics[width=1\linewidth]{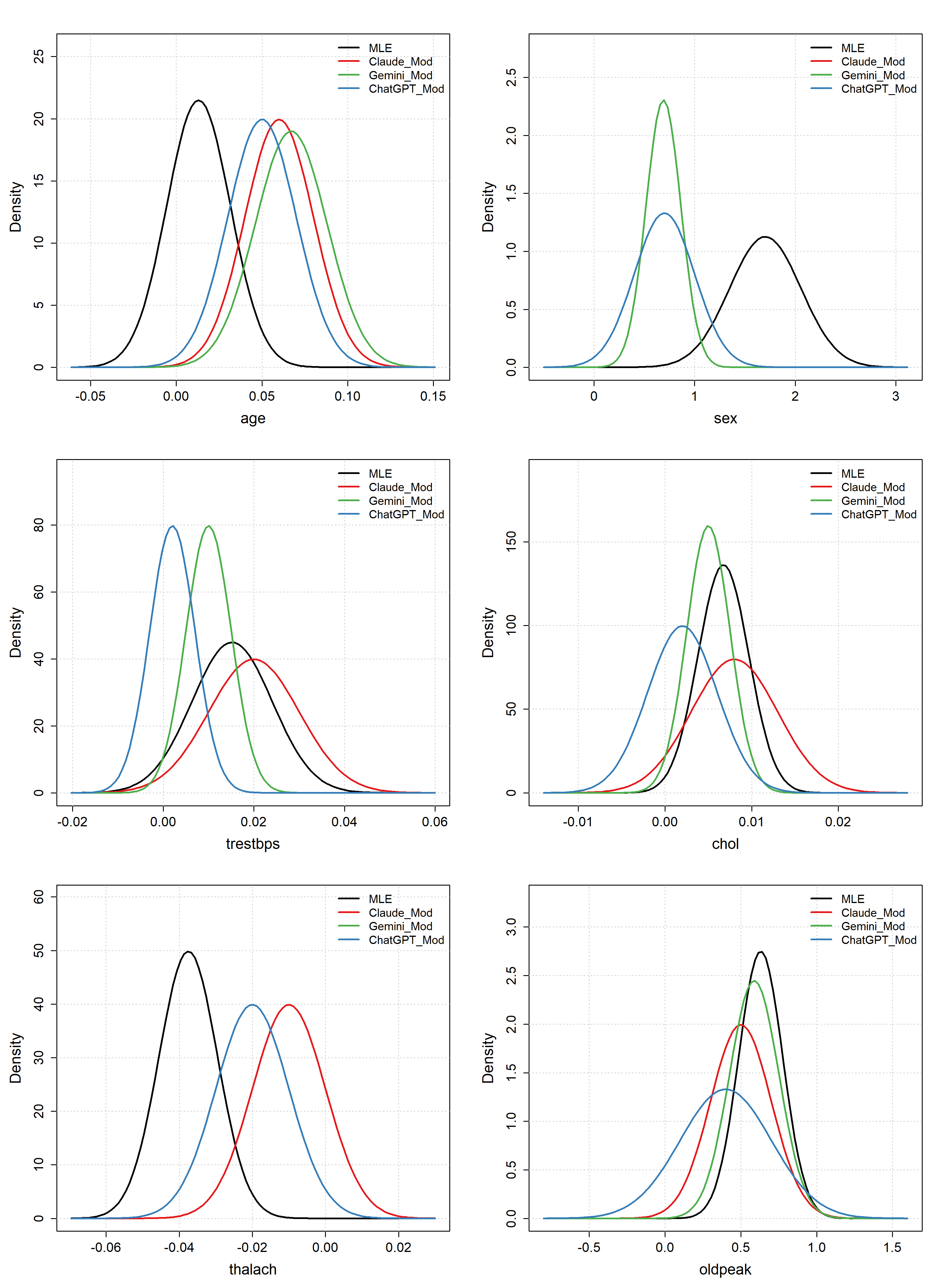}
\caption{Heart disease example: MLE distributions and the moderately informative LLM prior distributions for the different variables.}
\label{fig:1}
\end{figure}

Figure \ref{fig:1} shows the Gaussian approximation of the Maximum Likelihood Estimation (MLE) distribution and the suggested moderately informative priors for the three LLMs. Inspection of the plots reveals that in all cases, the prior distributions suggest the same sign for the log-odds ratio as the MLE. That is, if the data indicate that an increased value of a predictor increases the probability of CAD, the priors reflect the same directional belief. However, the expected magnitudes of the log-odds ratios can differ substantially. For the variable age, all three priors assume a far stronger association with CAD than what is observed in the data. This weak association is somewhat surprising, as it is well-established that the risk of CAD increases with age. The three prior distributions only partly overlap with the MLE, which indicates a significant discrepancy between the prior beliefs and the data. Similarly, for the variable sex, there is a disagreement, with the priors suggesting a lower log-odds ratio compared to the MLE. The narrowness of these priors suggests overconfidence from the LLMs, as the data are quite surprising relative to these distributions. For trestbps, Claude's prior aligns well with the MLE, while the other two LLMs suggest smaller log-odds ratios. For chol and oldpeak, the prior distributions appear to overlap well with the likelihood and show good agreement on the log-odds ratio. Finally, for thalach, the prior distributions suggest smaller log-odds ratios compared to the data. Furthermore, the priors are overly confident, resulting in poor coverage of the MLE distribution.

\begin{figure}
\centering
\includegraphics[width=1\linewidth]{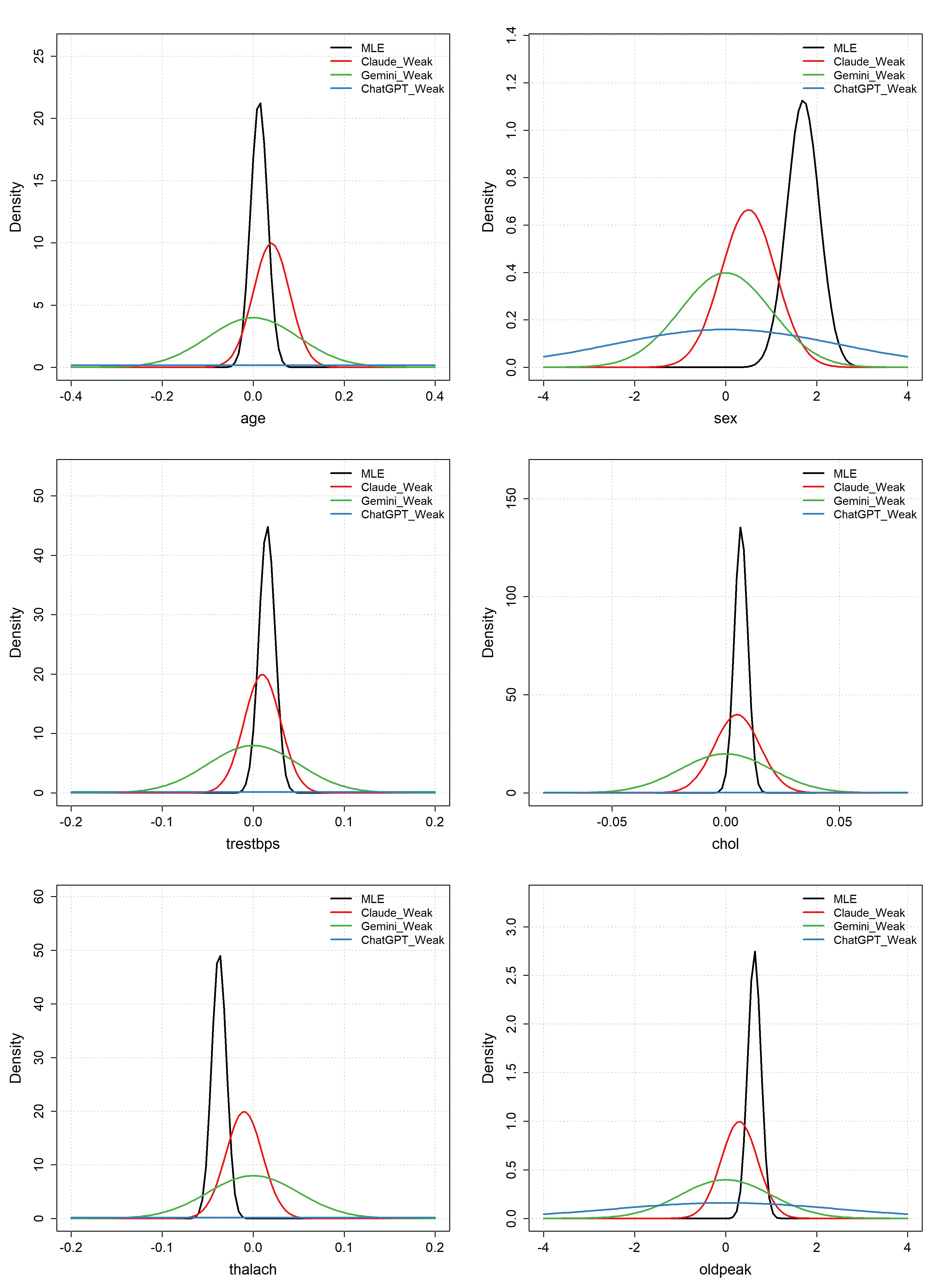}
\caption{Heart disease example: MLE distributions and the weakly informative LLM prior distributions for the different variables.}
\label{fig:2}
\end{figure}

Figure \ref{fig:2} shows the Gaussian approximation of the MLE distribution and the weakly informative priors suggested by the three LLMs. Comparing Figures \ref{fig:1} and \ref{fig:2} reveals that the suggested weakly informative priors are substantially wider than the moderately informative ones. For the variable sex, the prior distributions suggested by Gemini and Claude are still somewhat overconfident, resulting in fairly poor coverage of the MLE. Apart from this, the other suggested prior distributions appear to cover the MLE well. Furthermore, the priors suggested by ChatGPT are extremely wide, rendering them unnecessarily non-informative. A further disadvantage with the priors suggested by ChatGPT and Gemini is that the expected value of the prior distributions was set to 0, meaning that, a priori, they did not suggest any effect for the variables. We observed from the moderately informative priors that the LLMs had good knowledge about the effect of the different variables on CAD, so this seems unnecessarily vague. Claude, in contrast, maintained a non-zero expected effect in its priors while still achieving good coverage of the MLE.

Table \ref{tab:2} shows the computed Kullback-Leibler (KL) divergence between the MLE distribution and the different prior distributions. We see that Claude's weakly informative prior has the lowest average KL divergence and thus represents the best prior distribution according to this metric. This was the only prior set that was both informative (i.e., having an expected log-odds ratio different from zero) and not overly confident. Overall, Claude Opus and Gemini 2.5 Pro performed better than ChatGPT-4o-mini. As all three are top-performing models, it was not obvious beforehand which would perform best in these experiments. The LLMs had substantially more difficulty with some variables than others, with sex proving to be particularly challenging. The reason for this difficulty is not immediately clear. The weakly informative prior from ChatGPT performed poorly due to its excessive width. With the exception of ChatGPT, the weakly informative prior sets appear to be superior to the moderately informative ones, in which the LLMs were generally overconfident.
\begin{table}
\centering
\caption{Heart disease example: KL divergence between the MLE distribution and the different suggested prior sets. The first row shows which LLM suggested the prior sets and how informative they are (weak or moderate). The second to last row shows the average of the KL divergence for the different variables, and the last row shows the average ranking in terms of the best score (lower KL divergence is better).}
\label{tab:2}
\begin{tabular}{lcccccc}
\toprule
\textbf{LLM} & \textbf{Claude} & \textbf{Gemini} & \textbf{Claude} & \textbf{ChatGPT} & \textbf{Gemini} & \textbf{ChatGPT} \\
\textbf{Informative} & \textbf{Weak} & \textbf{Weak} & \textbf{Mod} & \textbf{Mod} & \textbf{Mod} & \textbf{Weak} \\
\midrule
age & 0.60 & 1.21 & 2.77 & 1.72 & 3.32 & 4.40 \\
sex & 2.21 & 2.05 & 5.61 & 5.61 & 18.00 & 1.70 \\
trestbps & 0.44 & 1.29 & 0.13 & 3.98 & 1.04 & 5.14 \\
chol & 0.79 & 1.49 & 0.24 & 0.78 & 0.27 & 6.25 \\
thalach & 1.44 & 1.63 & 3.83 & 1.58 & 1.58 & 5.24 \\
oldpeak & 0.92 & 1.64 & 0.30 & 0.64 & 0.05 & 2.38 \\
\textbf{Avg KL Div.} & \textbf{1.07} & \textbf{1.55} & \textbf{2.15} & \textbf{2.38} & \textbf{4.04} & \textbf{4.19} \\
\textbf{Avg Rank} & \textbf{2.50} & \textbf{3.67} & \textbf{2.83} & \textbf{3.33} & \textbf{3.17} & \textbf{5.17} \\
\bottomrule
\end{tabular}
\end{table}

Finally, we explored the prediction performance for the logistic Bayesian models using the different prior sets. The models were evaluated using five-fold cross-validation, and the performance was measured using the Brier score, mean negative log-score (MNLS), and AUC. We fitted the models using the R-INLA package~\cite{rue2009approximate}. The method approximates the posterior distribution using sophisticated integrated nested Laplace approximations. It is a specialized and robust alternative to the more general variational inference method and is also far more computationally efficient than commonly used Markov chain Monte Carlo methods. The results are shown in Table \ref{tab:4}, where the 'Frequentist' model refers to the logistic regression model fitted without the use of prior distributions. The results show some improvements in prediction performance for the Bayesian models using the LLM priors compared to the standard logistic regression model, but not statistically significant. P-values were computed using the Nadeau-Bengio corrected t-test~\cite{nadeau1999inference}. This is an important finding in itself and is consistent with Bayesian theory. Given the large size of the heart disease dataset, the information from the likelihood naturally dominates the prior, meaning significant performance gains from including prior information were not expected. This highlights that the real predictive utility of these priors is likely to be found in scenarios where prior information is more influential, such as in smaller datasets or for out-of-distribution generalization, representing an interesting direction for future research.
\begin{table}
\centering
\caption{Heart disease example: Prediction performance for the different Bayesian models and a frequentist logistic regression model. The values in the parentheses are p-values testing if the Bayesian models perform better than the frequentist model. The arrows indicate whether lower ($\downarrow$) or higher ($\uparrow$) values for the performance metrics are better.}
\begin{tabular}{llll}
\hline
\textbf{Model} & \textbf{Brier Score $\downarrow$} & \textbf{Mean Neg Log Score $\downarrow$} & \textbf{AUC $\uparrow$} \\
\hline
Frequentist & 0.1753 & 0.5281 & 0.8131 \\
Claude\_Mod & 0.1743 (p=0.488) & 0.5230 (p=0.477) & 0.8153 (p=0.380) \\
Claude\_Weak & 0.1737 (p=0.456) & 0.5219 (p=0.449) & 0.8158 (p=0.169) \\
Gemini\_Mod & 0.1737 (p=0.484) & 0.5207 (p=0.472) & 0.8156 (p=0.381) \\
Gemini\_Weak & 0.1743 (p=0.457) & 0.5244 (p=0.448) & 0.8147 (p=0.182) \\
ChatGPT\_Mod & 0.1739 (p=0.482) & 0.5209 (p=0.469) & 0.8165 (p=0.305) \\
ChatGPT\_Weak & 0.1749 (p=0.450) & 0.5262 (p=0.441) & 0.8135 (p=0.308) \\
\hline
\end{tabular}
\label{tab:4}
\end{table}

\subsection{Concrete Compressive Strength}
\label{sec:concrete}

In this example, we used a dataset from the UCI Machine Learning Repository~\cite{concrete_compressive_strength_165} to analyse the association between the variables in Table \ref{tab:3} and the compressive strength (CCS), measured in MPa, of the resulting concrete. The Age variable refers to how long the concrete was allowed to cure before the compressive strength test was performed. The other variables represent the quantity of each component in the concrete mixture. We model the association between CCS and the variables using a multiple linear regression model with Gaussian prior distributions for the regression coefficients.
\begin{table}
\centering 
\caption{Description of Concrete Components}
\label{tab:3}
\begin{tabular}{lll}
\toprule
\textbf{Variable description} & \textbf{Description} & \textbf{Measuring unit} \\
\midrule
Age & Curing time & day \\
Cement (component 1) & Binder & kg/m$^3$ \\
Blast Furnace Slag (component 2) & Cement replacement & kg/m$^3$ \\
Fly Ash (component 3) & Cement replacement & kg/m$^3$ \\
Water (component 4) & Activates cement & kg/m$^3$ \\
Superplasticizer (component 5) & Water reducer & kg/m$^3$ \\
Coarse Aggregate (component 6) & Filler, strength & kg/m$^3$ \\
Fine Aggregate (component 7) & Filler, workability & kg/m$^3$ \\
\bottomrule
\end{tabular}
\end{table}

\begin{figure}
\centering
\includegraphics[width=1\linewidth]{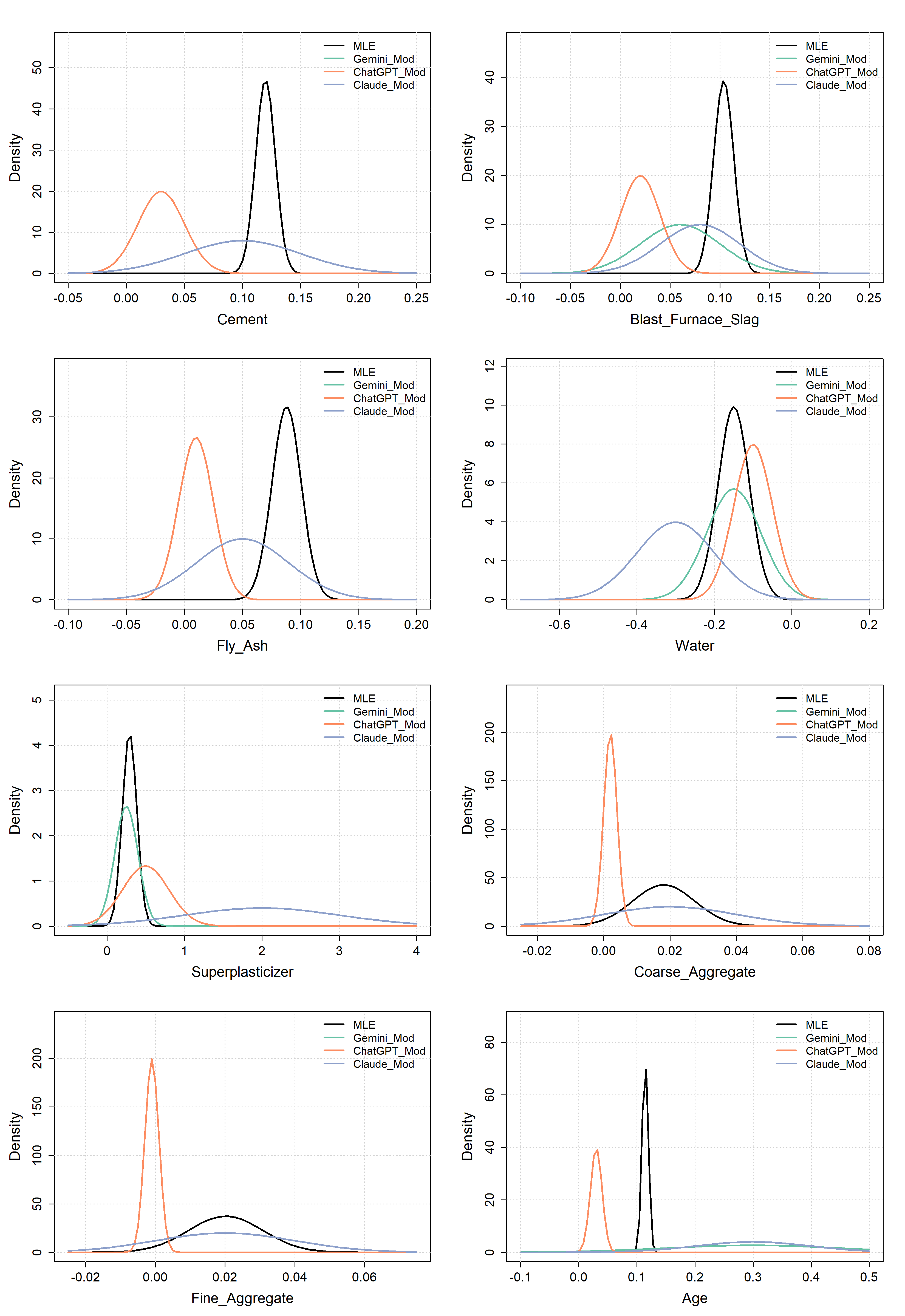}
\caption{Concrete strength example: MLE distributions and the moderately informative LLM prior distributions for the different variables.}
\label{fig:3}
\end{figure}

\begin{figure}
\centering
\includegraphics[width=1\linewidth]{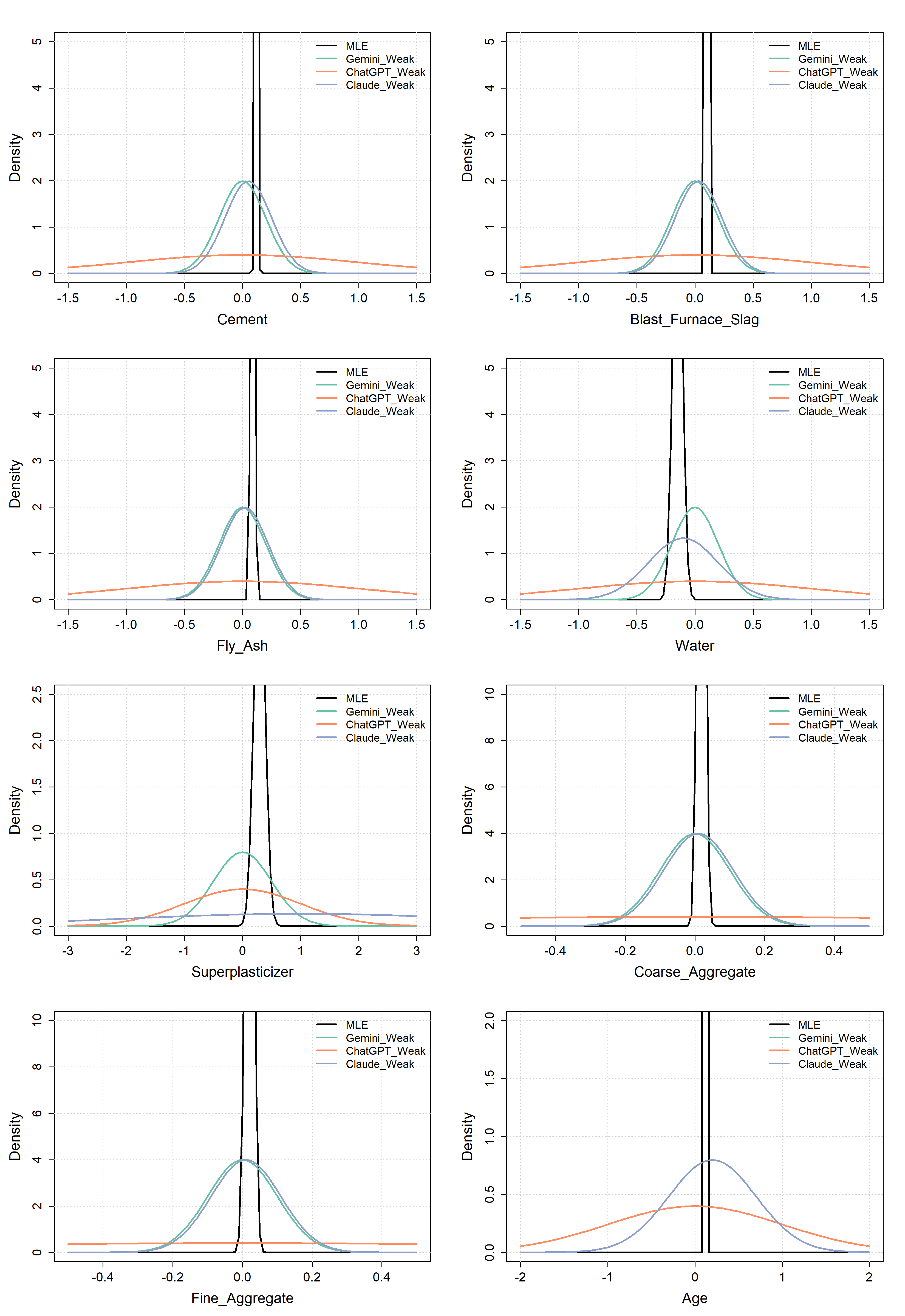}
\caption{Concrete strength example: MLE distributions and the weakly informative LLM prior distributions for the different variables.}
\label{fig:4}
\end{figure}

Figure \ref{fig:3} and \ref{fig:4} show the MLE distributions and the moderately and weakly informative prior distributions, respectively.

The findings from this example were largely consistent with those from the heart disease example. For the moderately informative priors, all LLMs correctly identified the direction of the effects, for instance, that CCS is positively associated with the quantity of cement and negatively associated with the quantity of water. While there was some disagreement on the magnitude of these effects, both Claude and Gemini adjusted the width of their prior distributions to avoid overconfidence. ChatGPT, on the other hand, was too confident for many of the variables. For the weakly informative priors, again, only Claude suggested priors that were simultaneously informative (i.e., with an expected effect different from zero) and weak. All LLMs provided wide prior distributions, and none of the weakly informative priors were overly confident. ChatGPT's priors were extremely wide, making them unnecessarily non-informative.

Table \ref{tab:5} shows the computed KL divergence between the MLE distribution and the different prior distributions. The moderately informative priors from both Gemini and Claude resulted in a low average KL divergence, indicating they were informative without being overconfident. All models struggled with the Age variable, generally defaulting to wide prior distributions. The association between CCS and age is highly non-linear; the strength increases rapidly during the first few days and weeks before gradually stabilizing. The prompt informed that the time span for the age variable was from one to 365 days, but it is possible that the LLMs struggled to effectively utilize this information, thus resorting to wider priors. ChatGPT generally struggled to provide suitable prior distributions, resulting in high KL divergence values.

\begin{table}
\centering
\caption{Concrete strength example: KL divergence between the MLE distribution and the different suggested prior sets. The first row shows which LLM suggested the prior sets and how informative they are (weak or moderate). The second to last row shows the average of the KL divergence for the different variables, and the last row shows the average ranking in terms of the best score (lower KL divergence is better).}
\label{tab:5}
\begin{tabular}{lcccccc}
\toprule
\textbf{LLM} & \textbf{Gemini} & \textbf{Claude} & \textbf{Gemini} & \textbf{Claude} & \textbf{ChatGPT} & \textbf{ChatGPT} \\
\textbf{Informative} & \textbf{Mod} & \textbf{Mod} & \textbf{Weak} & \textbf{Weak} & \textbf{Weak} & \textbf{Mod} \\
\midrule
Cement & 1.37 & 1.37 & 2.84 & 2.72 & 4.28 & 10.53 \\
Blast Furnace Slag & 1.51 & 1.08 & 2.62 & 2.55 & 4.10 & 9.10 \\
Fly Ash & 1.16 & 1.16 & 2.36 & 2.33 & 3.88 & 13.52 \\
Water & 0.22 & 1.62 & 1.41 & 1.53 & 2.73 & 0.54 \\
Superplasticizer & 0.21 & 3.33 & 1.37 & 3.00 & 1.92 & 0.95 \\
Coarse Aggregate & 0.37 & 0.37 & 1.89 & 1.87 & 4.17 & 41.33 \\
Fine Aggregate & 0.27 & 0.27 & 1.76 & 1.75 & 4.04 & 68.27 \\
Age & 3.59 & 4.14 & 4.72 & 4.04 & 4.72 & 35.73 \\
\textbf{Avg KL Div.} & \textbf{1.09} & \textbf{1.67} & \textbf{2.37} & \textbf{2.47} & \textbf{3.73} & \textbf{22.50} \\
\textbf{Avg Rank} & \textbf{1.12} & \textbf{2.38} & \textbf{3.75} & \textbf{3.25} & \textbf{4.88} & \textbf{5.00} \\
\bottomrule
\end{tabular}
\end{table}

Table \ref{tab:6} shows the prediction performance in a five-fold cross-validation experiment using the metrics MNLS, root mean squared error (RMSE), and mean absolute error (MAE). As with the heart disease example, the Bayesian models demonstrated slight but not statistically significant improvements in performance.  This result was expected, as the large sample size of the concrete dataset means the information from the data outweighs the contribution of the prior, a standard outcome in Bayesian analysis.
\begin{table}
\caption{Concrete strength example: Prediction performance for the different Bayesian models and a frequentist multiple regression model. The values in the parentheses are p-values testing if the Bayesian models perform better than the frequentist model. The arrows ($\downarrow$) indicate that lower values for the performance metrics are better.}
\label{tab:6}
\centering
\begin{tabular}{llll}
\hline
\textbf{Model} & \textbf{MNLS $\downarrow$} & \textbf{RMSE $\downarrow$} & \textbf{MAE $\downarrow$} \\
\hline
Frequentist & 3.770 & 10.492 & 8.293 \\
Gemini\_Mod & 3.768 (p=0.436) & 10.474 (p=0.435) & 8.289 (p=0.484) \\
Gemini\_Weak & 3.770 (p=0.453) & 10.489 (p=0.445) & 8.292 (p=0.459) \\
ChatGPT\_Mod & 3.780 (p=0.533) & 10.596 (p=0.535) & 8.496 (p=0.585) \\
ChatGPT\_Weak & 3.770 (p=0.503) & 10.492 (p=0.486) & 8.293 (p=0.485) \\
Claude\_Mod & 3.769 (p=0.471) & 10.480 (p=0.470) & 8.296 (p=0.509) \\
Claude\_Weak & 3.770 (p=0.469) & 10.490 (p=0.459) & 8.294 (p=0.518) \\
\hline
\end{tabular}
\end{table}

\section{Closing Remarks}

In this paper, we have analysed the approach of using LLMs to generate prior distributions for Bayesian statistical models. While \cite{capstick2024using} document improvement in prediction performance using priors based on LLM information, we didn't observe similar improvements in our experiments. However, the LLMs were generally capable of identifying the correct direction for the different associations, e.g., that the strength of concrete is reduced by the amount of water. However, the magnitude of the association often differed between the data and the prior expectations. This was particularly evident with the moderately informative priors for the heart disease example, where many of the suggested priors were overconfident. Among the weakly informative priors, only Claude was able to suggest priors that were informative (i.e., where the expected association was non-zero) while remaining suitably wide.

In conclusion, our findings indicate that LLMs show considerable potential as an efficient and objective tool for generating informative prior distributions. Particularly encouraging is their consistent ability to identify the correct directional nature of various associations. Nevertheless, a significant challenge remains in calibrating the width of these priors, as the LLMs demonstrated a tendency towards both overconfidence and underconfidence. It is interesting to reflect over why they in many cases were overconfident. Is it an inherent artifact of how LLMs are trained to provide confident-sounding answers? Could the "simulated literature review" part of the prompt encourage the models to recall strong, textbook associations without the necessary nuance for a specific dataset's context?

Some of the observed differences in effect sizes between the data and the LLM-generated priors could also be attributable to biases in the data. Although the data collection protocols for both datasets are considered reliable, the potential for sampling bias always exists. For instance, in the heart disease dataset, the proportion of individuals with heart disease is 46\%, a much higher prevalence than in the general population. This could lead to a case-control sampling bias, creating discrepancies between the data and the prior distributions informed by population-level knowledge. Whether the disagreements between the data and the priors stem from biases in the data or from incorrect assessments by the LLMs is difficult to ascertain. However, for this analysis, we have proceeded under the assumption that the data are not severely biased, and we have evaluated the priors based on the criterion that they should not be overly surprised by the data.

There are several directions for future work. The prompts used in this study could likely be improved to increase the quality of the suggested priors. It is also interesting to explore this approach for other Bayesian models, such as latent variable models, hierarchical models, and graphical/causal models. Finally, exploring the potential of LLMs in out-of-distribution generalization, where the test data differs systematically from the training data, is a promising avenue. In such scenarios, LLM-generated priors could potentially serve as a bridge between the two data distributions, improving model robustness and performance.

\clearpage
\bibliographystyle{plain}
\bibliography{bibl}
\end{document}